\documentclass[conference]{IEEEtran}
\usepackage[utf8]{inputenc}
\usepackage{cite}
\usepackage{soul}
\usepackage{multirow}
\usepackage[table,xcdraw]{xcolor}

\ifCLASSINFOpdf
\usepackage[pdftex]{graphicx}
\graphicspath{{../pdf/}{../jpeg/}}
\DeclareGraphicsExtensions{.pdf,.jpeg,.png}
\else
\fi

\usepackage[justification=centering]{caption}

\usepackage{todonotes}
\usepackage[hidelinks]{hyperref}

\usepackage{amsmath}
\usepackage{array}
\newcolumntype{P}[1]{>{\centering\arraybackslash}p{#1}}

\ifCLASSOPTIONcompsoc
\usepackage{caption}
\fi

\usepackage[numbers]{natbib}

\usepackage{comment}
\usepackage{blindtext}
\usepackage{longtable}
\usepackage{lipsum} 
\usepackage{cite}
\usepackage{soul}
\usepackage{multirow}
\usepackage[table,xcdraw]{xcolor}

\ifCLASSINFOpdf
\usepackage[pdftex]{graphicx}
\renewcommand{\baselinestretch}{0.94}
\graphicspath{{../pdf/}{../jpeg/}}
\DeclareGraphicsExtensions{.pdf,.jpeg,.png}
\else
\fi
\usepackage[justification=centering]{caption}

\usepackage[savewrites,nopostdot,toc,acronym,symbols,nogroupskip]{glossaries-extra}
\setabbreviationstyle[acronym]{long-short}
\definecolor{added}{rgb}{0.9,0.9,1}

\usepackage{footnote}
\makesavenoteenv{tabular}
\makesavenoteenv{table}
\usepackage{booktabs, caption, makecell}

\usepackage{threeparttable}
\usepackage[export]{adjustbox}
\usepackage{algorithm}
\usepackage[noend]{algpseudocode}

\newcommand{\sfunction}[1]{\textsf{\textsc{#1}}}
\algrenewcommand\algorithmicforall{\textbf{foreach}}
\algrenewcommand\algorithmicindent{.8em}

\usepackage{glossaries}
\makenoidxglossaries
\newacronym{mmtc}{mMTC}{massive Machine Type Communication}
\newacronym{urllc}{uRLLC}{ultra-Reliable Low Latency Communication}
\newacronym{embb}{eMMB}{enhanced Mobile Broadband}
\newacronym{mec}{MEC}{Multi-Access Edge Computing}
\newacronym{bsaf}{BSAF}{Back-situation Awareness Function}
\newacronym{clm}{CLM}{Cooperative Lane Merge}
\newacronym{cam}{CAM}{Connected and Automated Mobility}
\newacronym{nfv}{NFV}{Network Function Virtualization}
\newacronym{mno}{MNO}{Mobile Network Operator}
\newacronym{3gpp}{3GPP}{$3{^{rd}}$ Generation Partnership Project}
\newacronym{v2x}{V2X}{Vehicle-to-Everything}
\newacronym{v2i}{V2I}{Vehicle-to-Infrastructure}
\newacronym{v2v}{V2V}{Vehicle-to-Vehicle}
\newacronym{v2n}{V2N}{Vehicle-to-Network}
\newacronym{dsrc}{DSRC}{Dedicated Short Range Communication}
\newacronym{rsu}{RSU}{Road Side Unit}
\newacronym{d2d}{D2D}{Device-to-Device}
\newacronym{nr}{NR}{New Radio}
\newacronym{5gaa}{5GAA}{5G Automotive Association}
\newacronym{rnis}{RNIS}{Radio Network Information Service}
\newacronym{ci/cd}{CI/CD}{Continuous Integration and Continuous Delivery}
\newacronym{mla}{MLA}{Management Level Agreement}
\newacronym{onap}{ONAP}{Open Network Automation Platform}
\newacronym{osm}{OSM}{Open Source MANO}
\newacronym{mano}{MANO}{Management and Orchestration}
\newacronym{rest}{REST}{REpresentational State Transfer}
\newacronym{nfv-lo}{NFV-LO}{NFV Local Orchestrator}
\newacronym{nfv-so}{NFV-SO}{NFV Service Orchestrator}
\newacronym{meao}{MEAO}{MEC Application Orchestrator}
\newacronym{fdio}{FDIO}{Fast Data Input Output}
\newacronym{lcm}{LCM}{Life-cycle Management}
\newacronym{gRPC}{gRPC}{Remote Procedure Call}
\newacronym{vas}{VAS}{Value-added Service}
\newacronym{sae}{SAE}{Society of Automotive Engineers}
\newacronym{nf}{NF}{Network Function}
\newacronym{vnf}{VNF}{Virtualized Network Function}
\newacronym{ssc}{SSC}{Service and Session Continuity}
\newacronym{kpi}{KPI}{Key Performance Indicator}
\newacronym{fa}{FA}{Federation Agent}
\newacronym{fm}{FM}{Federation Manager}
\newacronym{vim}{VIM}{Virtualized Infrastructure Manager}
\newacronym{vnfm}{VNFM}{VNF Manager}
\newacronym{upf}{UPF}{User Plane Function}
\newacronym{qos}{QoS}{Quality of Service}
\newacronym{pdn}{PDN}{Packet Data Network}
\newacronym{af}{AF}{Application Function}
\newacronym{kubernetes}{k8s}{Kubernetes}
\newacronym{etsi}{ETSI}{European Telecommunications Standards Institute}
\newacronym{sdn}{SDN}{Software Defined Networking}
\newacronym{gpcu}{GPCU}{General Purpose Computing Unit}
\newacronym{its-g5}{ITS-G5}{Intelligent Transportation System}
\newacronym{nfvi}{NFVI}{NFV Infrastructure}
\newacronym{cdn}{CDN}{Content Delivery Network}
\newacronym{cdnaas}{CDNaaS}{CDN as a Service}
\newacronym{cam1}{CAM}{Cooperative Awareness Message}
\newacronym{denm}{DENM}{Decentralized Environmental Notification Message}
\newacronym{api}{API}{Application Programmable Interface}
\newacronym{k8s}{Kubernetes}{Kubernetes}
\newacronym{cmeao}{cMEAO}{Centralized MEC Application Orchestrator}
\newacronym{lstm}{LSTM}{Long Short-Term Memory}
\newacronym{mcdm}{MCDM}{Multi-Criteria Decision Making}
\newacronym{poc}{PoC}{Poof-of-Concept}
\newacronym{ladn}{LADN}{Local Area Data Network}
\newacronym{eas}{EAS}{Edge Application Server}
\newacronym{eac}{EAC}{Edge Application Client}
\newacronym{eec}{EEC}{Edge Enabler Client}
\newacronym{ees}{EES}{Edge Enabler Server}
\newacronym{ecs}{ECS}{Edge Configuration Server}
\newacronym{rtt}{RTT}{Round Trip Time} 

\usepackage{float}
\usepackage{subfig}

\renewcommand{\baselinestretch}{0.94}

\begin{document}

\title{An optimized application-context relocation approach for Connected and Automated Mobility (CAM)}

\author{Nina Slamnik-Kriještorac, Steven Latré, and Johann M. Marquez-Barja

\\ \vspace{1mm}
University of Antwerp - imec, IDLab - Faculty of Applied Engineering, Belgium\\ 
E-mail: \{Nina.SlamnikKrijestorac, Steven.Latre, Johann.Marquez-Barja\}@uantwerpen.be \\
}

\maketitle
%
\begin{abstract}
In this paper, we study and present a management and orchestration framework for vehicular communications, which enables service continuity for the vehicle via an optimized application-context relocation approach. To optimize the transfer of the application-context for Connected and Automated Mobility (CAM) services, our MEC orchestrator performs prediction of resource availability in the edge infrastructure based on the Long Short-Term Memory (LSTM) model, and it makes a final decision on relocation by calculating the outcome of a Multi-Criteria Decision-Making (MCDM) algorithm, taking into account the i) resource prediction, ii) latency and bandwidth on the communication links, and iii) geographical locations of the vehicle and edge hosts in the network infrastructure. Furthermore, we have built a proof-of-concept for the orchestration framework in a real-life distributed testbed environment, to showcase the efficiency in optimizing the edge host selection and application-context relocation towards achieving continuity of a service that informs vehicle about the driving conditions on the road.  
\end{abstract}
\section{Introduction and Background} \label{sec:intro}
\begin{figure}[t!]
  \centering
  \includegraphics[width=90mm, height =75mm]{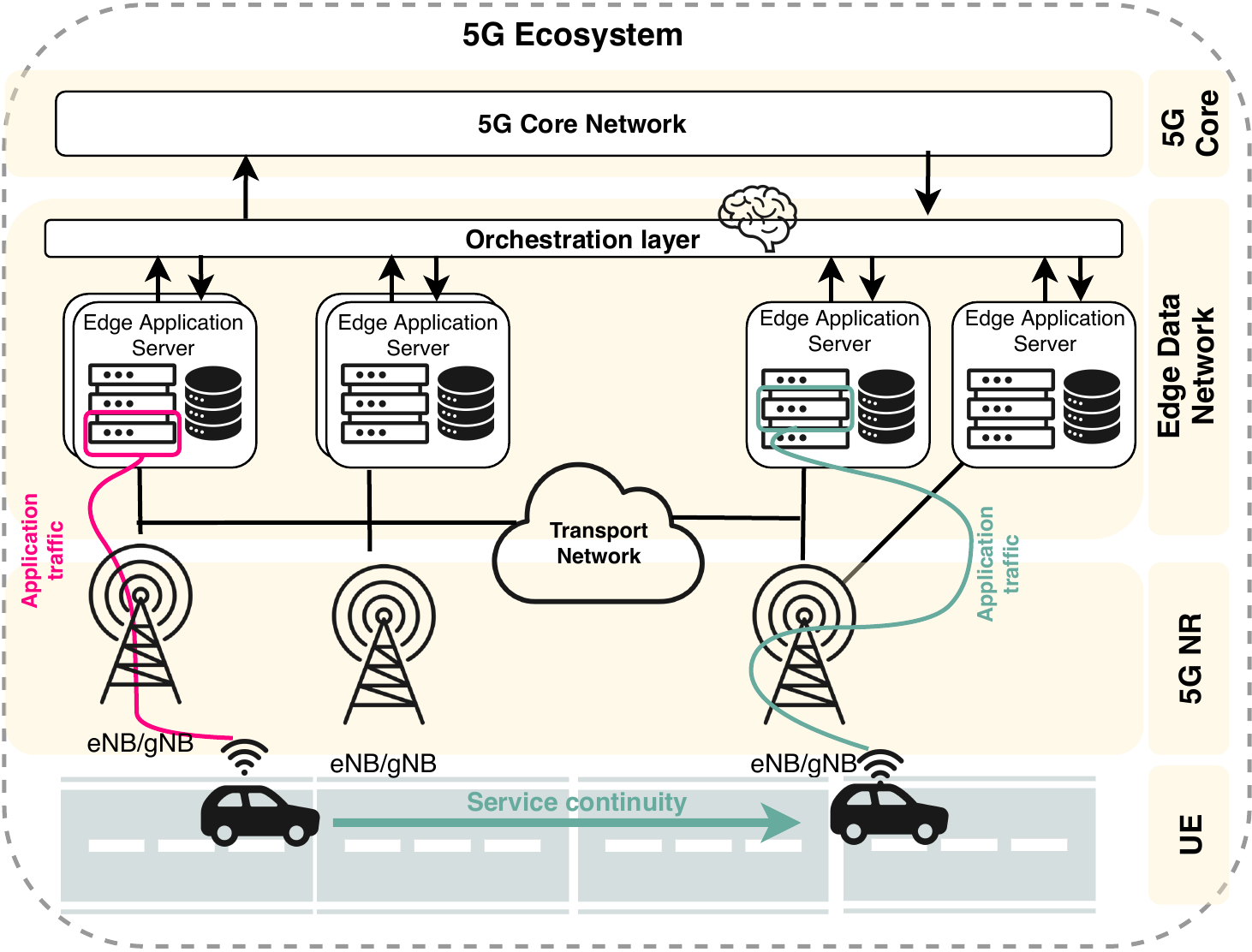}
  \caption{Enabling CAM service continuity for vehicles in 5G ecosystem.}
  \label{fig:fig1}
\end{figure}

The 5G ecosystem illustrated in Fig. \ref{fig:fig1}, consists of the 5G Core and \gls{nr}, including the managed and orchestrated distributed edge network infrastructure. In such ecosystem, a vehicle is capable to collect the contextual driving information, thereby connecting to the \gls{cam} services, located at the edge in order to keep the communication latency to a minimum possible level. In particular, to be less dependent on driver's actions, and to ensure higher safety, the vehicle needs to receive instructions from the network infrastructure in less than 100ms \cite{Zhang2020}, which requires service availability close to the vehicles, i.e., in the edge infrastructure such as \gls{mec} platforms, as well as transferring the application traffic via 5G Uu interface \cite{Kanavos2021}. Thanks to the \gls{sdn} and \gls{nfv}, \gls{mec} platforms can offer distributed cloud-native service deployments at a closer proximity, enhancing the user experience and increasing network performance. However, due to the high mobility of users, such distributed service deployment requires an agile reconfiguration and constant monitoring in order to maintain the service continuity. 
Thus, in this paper, we present a management and orchestration framework that enables service continuity in a highly mobile environment, with the reference to the \gls{3gpp} architecture for enabling edge applications \cite{3gppedgeapp}, and ETSI NFV MANO framework \cite{etsinfv}. The service continuity is enabled via an optimized application-context relocation approach that is triggered by a MEC application orchestrator while a vehicle, which is a consumer of the \gls{cam} service on the edge, moves along the road. 

To efficiently solve the challenges on how and when to perform application-context relocation, the MEC orchestrator in our framework is performing the prediction of resource availability in edge \gls{nfvi}, utilizing the prediction model based on \gls{lstm} \cite{Violos2020}, and making a decision on the optimal application service placement by running the Technique for Order Preference by Similarity to Ideal Solution (TOPSIS) algorithm, i.e., one of the widely adopted \gls{mcdm} concepts \cite{Singh2014}, thereby taking into account: i) the aforementioned resource availability prediction, ii) the latency and bandwidth on the communication path to the vehicle, and iii) geographical locations of vehicle and \gls{mec} host in the edge infrastructure. 
To measure the performance of the \gls{mec} application orchestrator, we have built a \gls{poc} of the management and orchestration framework in a real-life distributed testbed environment, combining the Virtual Wall\footnote{Virtual Wall: \url{https://doc.ilabt.imec.be/ilabt/virtualwall/}} testbed, and the Smart Highway\footnote{Smart Highway: \url{https://www.fed4fire.eu/testbeds/smart-highway/}} testbed, both located in Belgium. 

\textcolor{black}{Since the autonomous vehicles need to continuously collect the data from surrounding environment and network infrastructure, including the suggestions on braking and accelerating without driver assistance, the experimentation in our PoC reflects such a use case in which MEC application service is informing vehicle about driving conditions on the road (e.g., traffic jams, poor weather conditions, emergency situations, etc.). Thanks to the distributed service deployment, vehicle is being informed about driving conditions not only in its close proximity, but also in extended regions, thereby enabling vehicle to choose another route for its manoeuvre.}
The \gls{poc} is further described in Section \ref{sec:poc}, where we also show the improvement in the application server response time when application-context relocation is performed, thereby proving the efficiency of the \gls{mec} application orchestrator in optimizing the \gls{mec} host selection and application-context relocation towards achieving service continuity. 
\section{Application-context relocation} \label{sec:relocation}
\begin{figure}[t!]
  \centering
  \includegraphics[width=0.85\columnwidth]{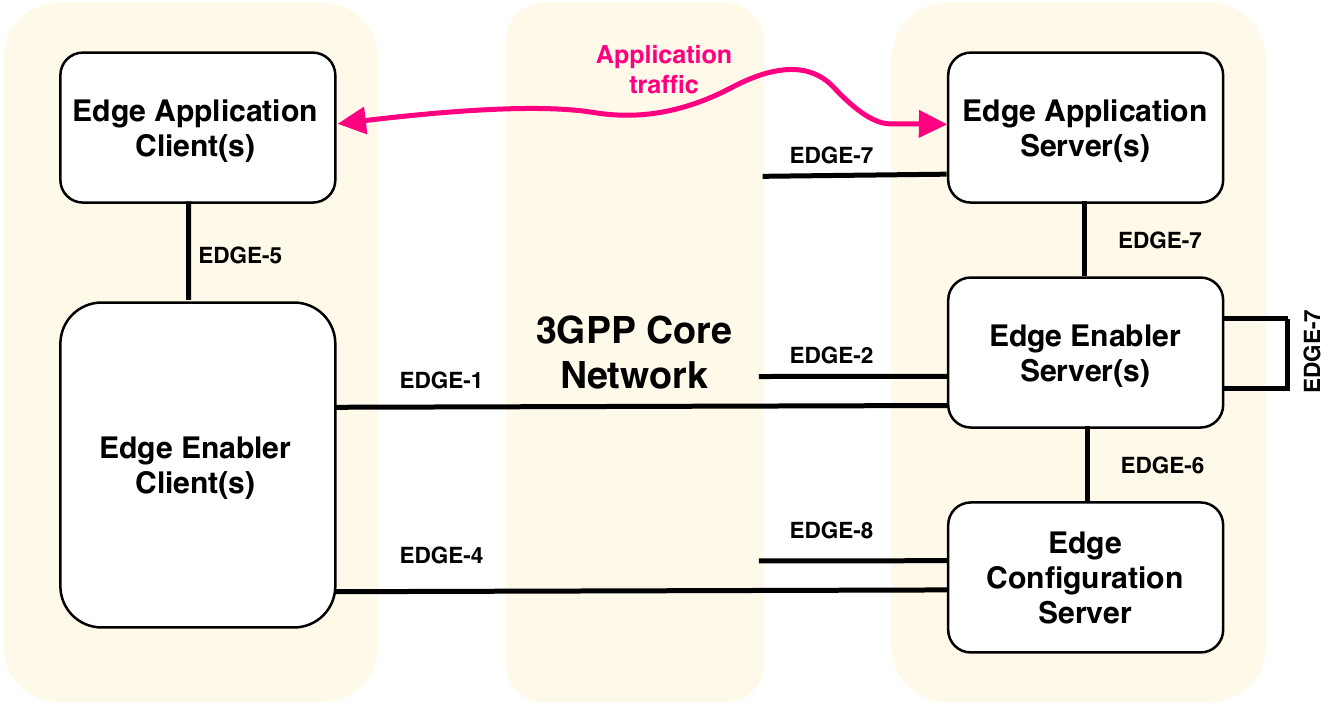}
  \caption{3GPP Architecture for Enabling Edge Applications.}
  \label{fig:3gpp}
\end{figure}
\subsection{Management and Orchestration framework}
As a part of Release 17, \gls{3gpp} is standardizing an architecture for enabling edge applications, while providing mutual awareness between edge client applications (i.e., in-vehicle application), and edge application servers running in the edge data network. This \gls{3gpp} standardization track \cite{3gppedgeapp} created i) the application layer architecture, which is shown in Fig. \ref{fig:3gpp}, ii) procedures, and iii) information flows necessary for enabling edge applications over \gls{3gpp} networks. 
In particular, in architecture shown in Fig. \ref{fig:3gpp}, the edge network consists of i) \gls{ecs}, which provides configuration data, i.e., \gls{ladn} URI, to the \gls{eec} to connect to the \gls{ees}, ii) \gls{ees}, which interacts with \gls{3gpp} core to collect network and service capabilities (e.g., location services, \gls{qos} management, etc.) that will improve the performance of edge application server, thereby enabling \gls{eac} to connect to the server, and iii) \gls{eas}, which performs server functions and exchanges application data traffic with the client (Figures \ref{fig:fig1} and \ref{fig:3gpp}). On the client side, in our case in the vehicle, \gls{eec} discovers the edge network, retrieves the necessary information for connecting to the edge (e.g., coverage area/service area, types of application servers or \gls{mec} applications, etc.), and connects to it via IP address provided by \gls{ees}. Furthermore, different reference points, i.e., EDGE 1-EDGE 7, are defined to enable communication between different architecture elements.

\begin{figure}[t!]
  \centering
  \includegraphics[width=\columnwidth]{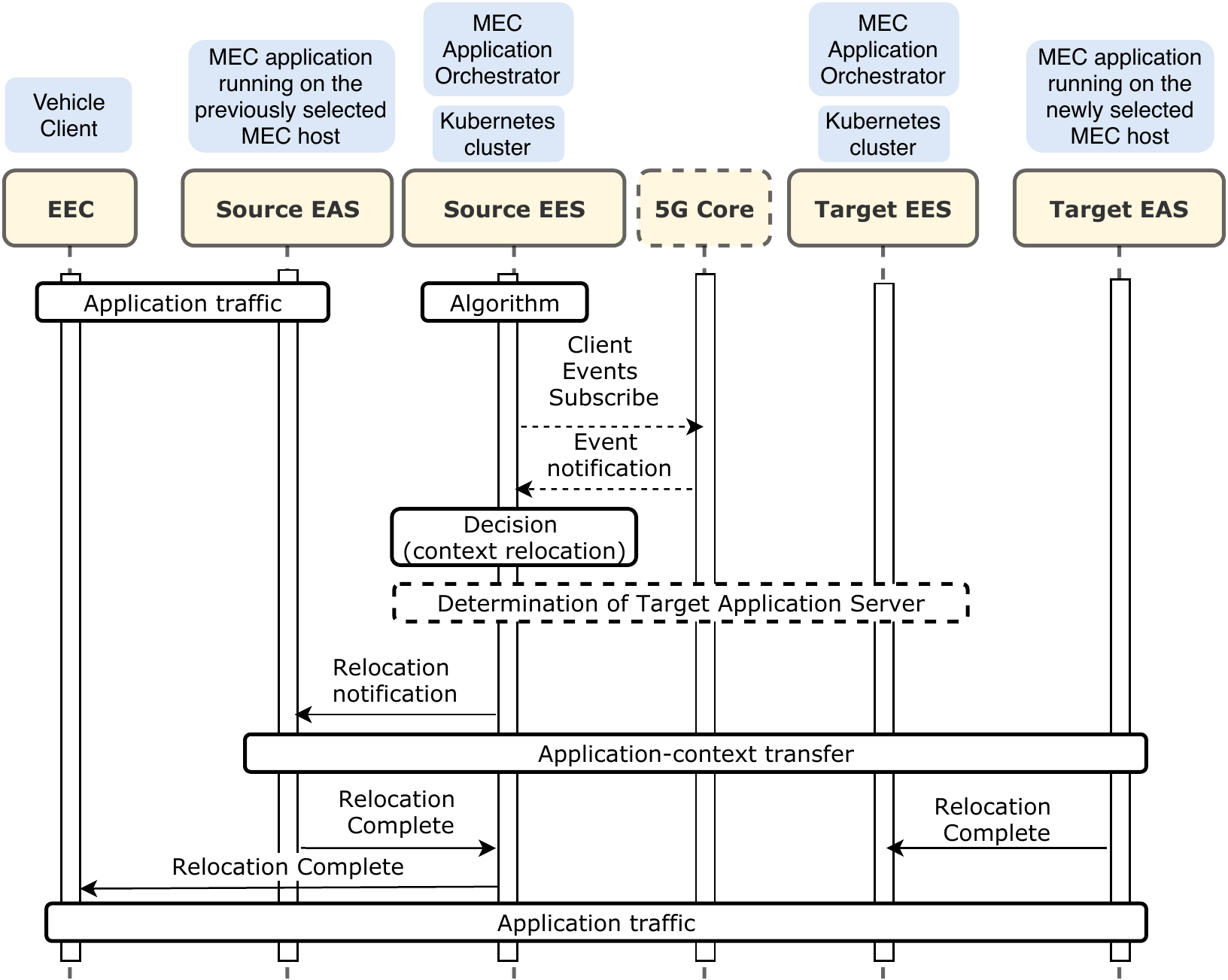}
  \caption{Message Sequence Chart for the application-context relocation procedure.}
  \label{fig:msc}
\end{figure}

In Fig. \ref{fig:msc}, we present the message sequence chart to showcase the operation of the application-context relocation from one edge to another, thereby mapping our management and orchestration framework (black boxes on the top), which is based on \gls{etsi} \gls{nfv} \gls{mano} \cite{etsinfv} and presented in \cite{ccnc}, to the 3GPP architecture for enabling edge applications (yellow boxes). In particular, when vehicle sends a discovery request to the \gls{mec} orchestrator, as a response, it receives a list of all available \gls{mec} application services that corresponds to the filters applied in the request. This way, the vehicle becomes edge-aware, as it can connect to any application server from the list. Once MEC orchestrator decides that vehicle needs to connect to another MEC application service due to e.g., increased resource consumption that will degrade the \gls{qos}, vehicle going out of the geographical service area, vehicle re-attaching from one \gls{upf} anchor to another, etc., the same reference point, i.e., EDGE-1, is used to inform vehicle about the newly selected MEC host (i.e., Relocation complete notification in Fig. \ref{fig:msc}). Furthermore, this notification contains the endpoint of the new \gls{mec} application instance running on the new \gls{mec} host, and client in the vehicle needs to be configured in the way that it can dynamically change the IP endpoint of the application server from which it consumes the service. %
%
%
%
%
\subsection{Optimized MEC host selection}
To transfer the context of application service that vehicle is consuming, and to enable this vehicle to continue utilizing the service in a seamless way, we need to i) identify a corresponding target \gls{mec} host, ii) perform transfer of application-context, iii) reconfigure the traffic rules and management policies, and iv) setup a new communication path to the vehicle. The step i) is performed by our MEC application orchestrator that is designed as an extension of Kubernetes master role. It runs the optimized \gls{mec} host selection algorithm, thereby predicting the resource availability in all \gls{mec} hosts that belong to the management and orchestration framework, by applying the \gls{lstm} based prediction. Furthermore, taking into account the predicted resource availability, the latency and bandwidth on the communication path to the vehicle, and geographical location of both vehicle and \gls{mec} hosts, the orchestrator makes decision whether application-context needs to be transferred to another edge or not, by performing the \gls{mcdm} analysis. If the decision is made, and new node is selected for application placement, orchestrator instantiates new application service on the target \gls{mec} host, and allows application services from the source host to transfer the context to the target host, as shown in Fig. \ref{fig:msc}. Finally, once the context is transferred, the orchestrator sends a notification to the edge-aware client application in the vehicle, which then starts consuming service from the new \gls{mec} host, after the traffic rules and management policies are reconfigured by the MEC application orchestrator. Due to the limited space in this manuscript, the detailed presentation of the prediction and MCDM algorithms is not included in this paper, but it will be part of an extended version.

%

\section{Proof of Concept} \label{sec:poc}
\begin{figure*}[!tbp]
	\setlength\belowcaptionskip{-2ex}
	\centering
	\subfloat[PoC setup.]{\label{fig:poc}\includegraphics[width=0.65\columnwidth]{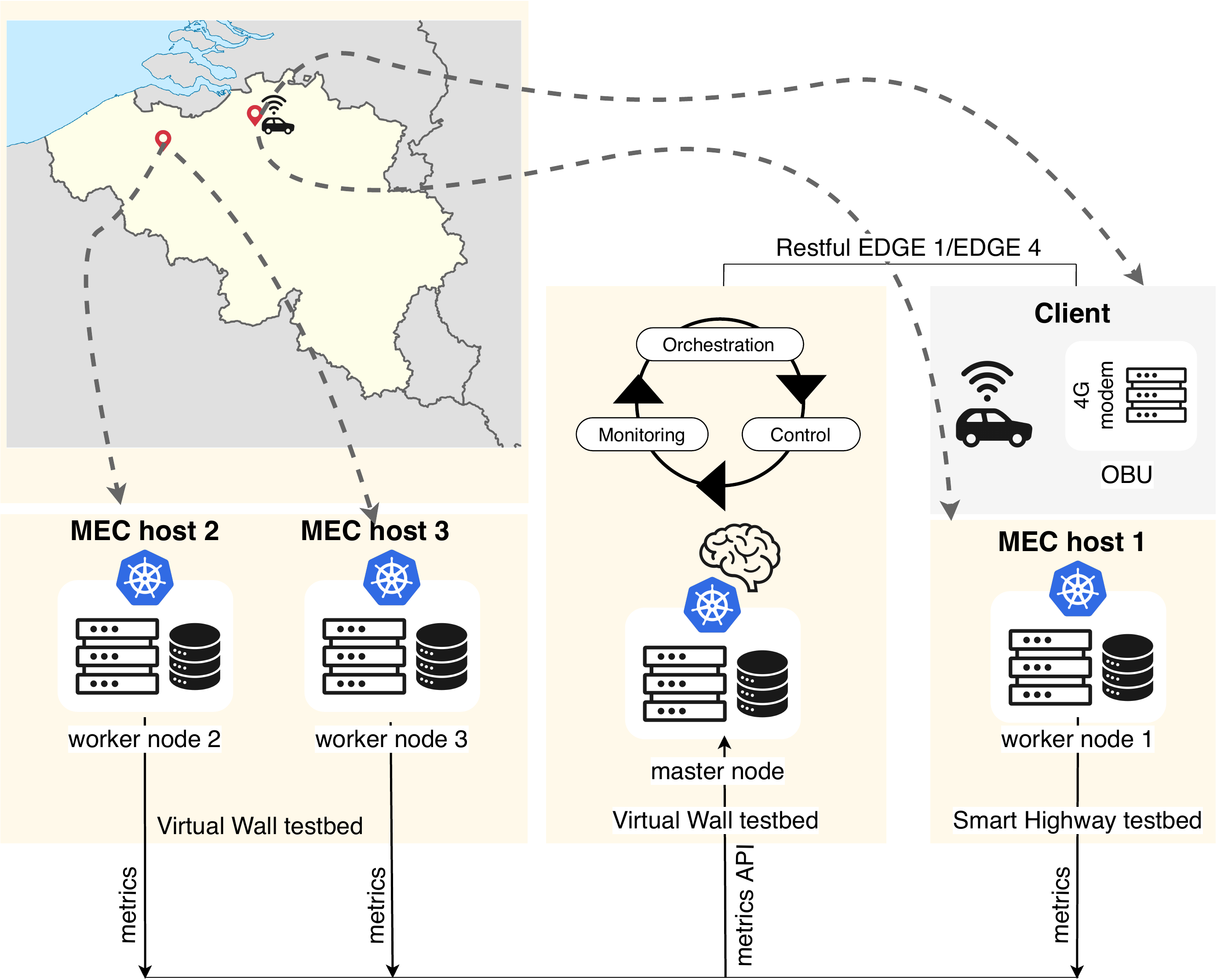}}\quad
	\subfloat[RTT.]{\label{fig:rtt}\includegraphics[width=0.65\columnwidth]{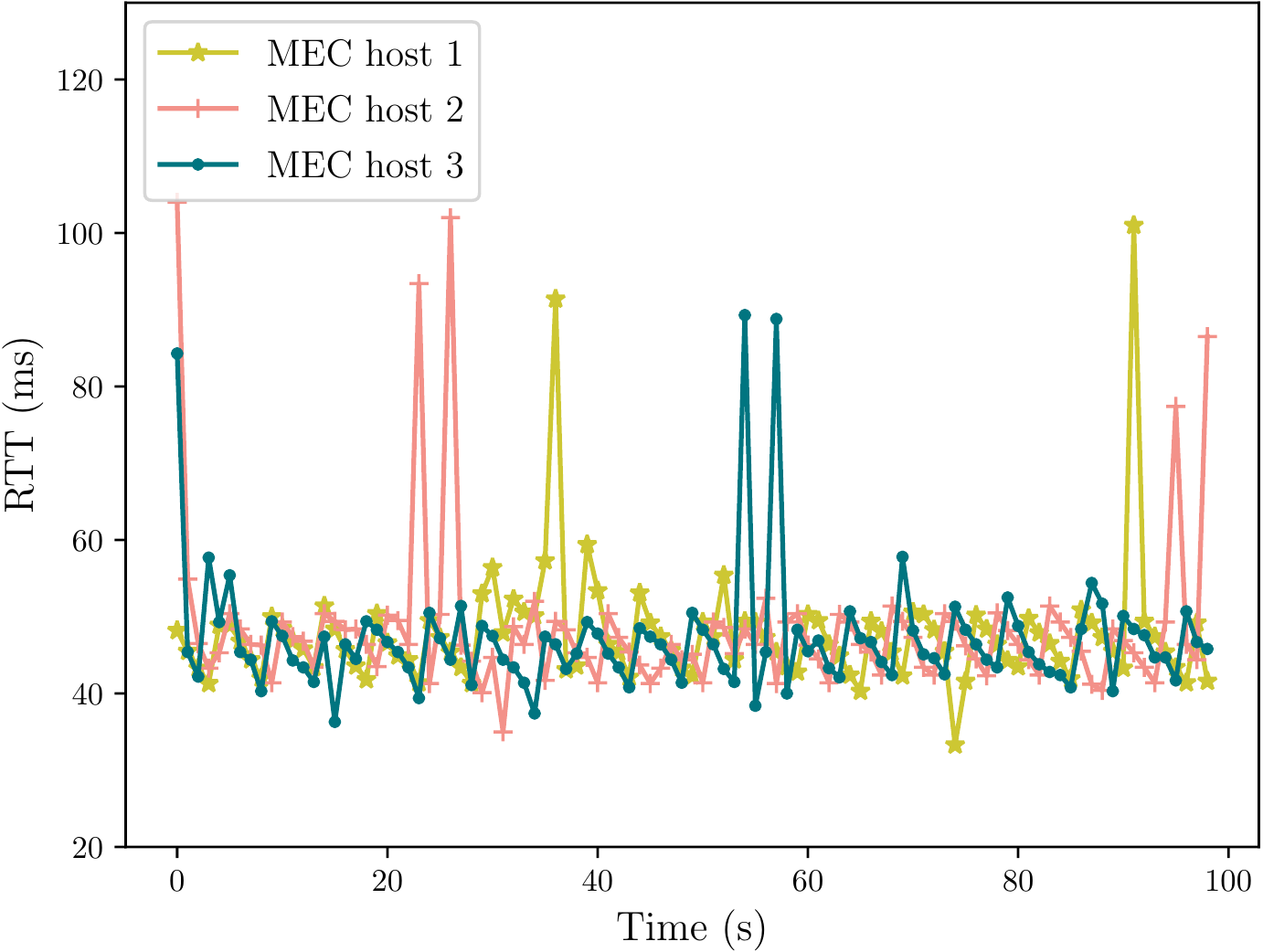}}\quad
	\subfloat[Response time.]{\label{fig:response}\includegraphics[width=0.65\columnwidth]{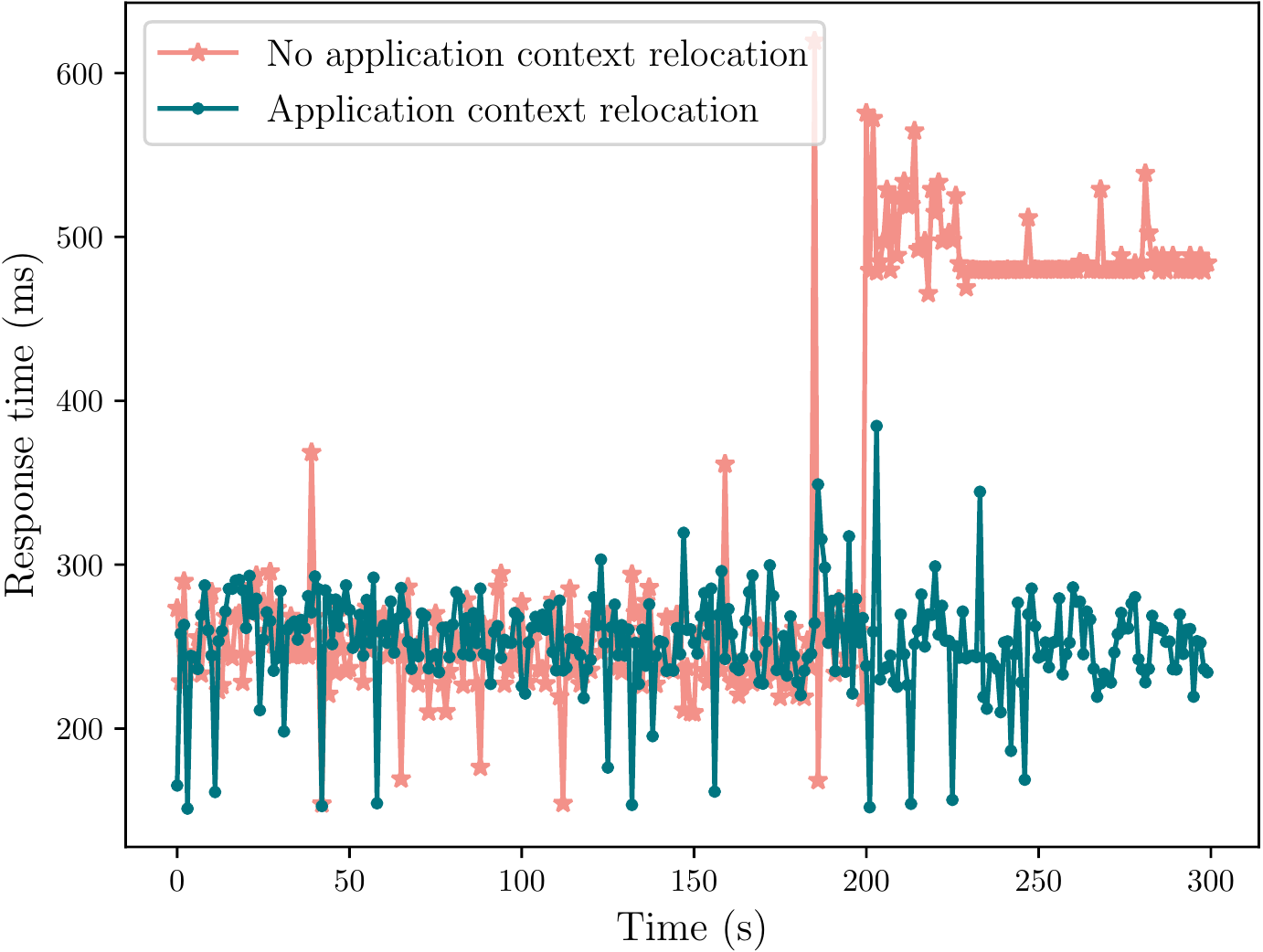}}\quad
	\caption{PoC in a distributed testbed environment.}\label{fig:pocoverall}
\end{figure*}
\subsection{Experimentation setup}
The \gls{poc} that we have built to measure the performance of our MEC application orchestrator while performing application-context relocation (described in Fig. \ref{fig:msc}) is illustrated in Fig. \ref{fig:poc}. The experimentation setup combines the components from two testbed facilities, the Virtual Wall testbed (Ghent, Belgium), which is a testbed for large networking and cloud experiments, and the Smart Highway testbed (Antwerp, Belgium), a test site built on top of the E313 highway for the purpose of \gls{v2x} research. 

The detailed specification of testbed machines of MEC hosts 1, 2, and 3, is presented in Table \ref{tab:poc}. The \gls{mec} hosts are utilized as distributed edge cloud environment where \gls{mec} application services are deployed, and the vehicle on-board unit is used as a client. The overall deployment is created in the Kubernetes\footnote{Kubernetes: \url{https://kubernetes.io/docs/home/}} cluster, in which the Kubernetes master is deployed on a separate bare-metal server with the same characteristics as the other two Virtual Wall hosts that are used as worker nodes. Our MEC application orchestrator is running on the master node to extend the capabilities of Kubernetes master towards supporting optimized MEC host selection for application-context relocation. Thus, the \gls{nfvi} in our \gls{poc} consists of three distributed \gls{mec} hosts, i.e., two bare-metal servers in Virtual Wall, and one GPCU inside the \gls{rsu} that is located on the highway site, all three running as worker nodes in the same Kubernetes cluster. Furthermore, we have enabled the Metrics API in Kubernetes cluster to collect CPU, memory, and storage consumption from all distributed worker nodes, in order to train and validate our prediction algorithm.
The client in our \gls{poc} is deployed as a Docker-based web application, which is on-boarded within NUC in the on-board unit, and it is connected to the distributed \gls{mec} application services via long-range 4G. 
\begin{table}[]
\centering
\caption{System characteristics of the testbed machines.}
\begin{tabular}{|l|l|l|l|l|}
\hline
\multicolumn{5}{|c|}{\textbf{PoC information}} \\ \hline
\textbf{Type} & \textbf{\begin{tabular}[c]{@{}l@{}}MEC \\ host 1\\ RSU\end{tabular}} & \multicolumn{1}{c|}{\textbf{\begin{tabular}[c]{@{}c@{}}MEC \\ host 2\end{tabular}}} & \multicolumn{1}{c|}{\textbf{\begin{tabular}[c]{@{}c@{}}MEC \\ host 3\end{tabular}}} & \multicolumn{1}{c|}{\textbf{\begin{tabular}[c]{@{}c@{}}Vehicle\\ NUC\end{tabular}}} \\ \hline
Testbed & \begin{tabular}[c]{@{}l@{}}Smart \\ Highway\end{tabular} & \begin{tabular}[c]{@{}l@{}}Virtual \\ Wall\end{tabular} & \begin{tabular}[c]{@{}l@{}}Virtual \\ Wall\end{tabular} & \begin{tabular}[c]{@{}l@{}}Smart \\ Highway\end{tabular} \\ \hline
Location & Antwerp & Ghent & Ghent & Antwerp \\ \hline
\begin{tabular}[c]{@{}l@{}}CPU\\ (GHz)\end{tabular} & 1.280 & 2.252 & 2.252 & 1.9 \\ \hline
\begin{tabular}[c]{@{}l@{}}RAM \\ (GB)\end{tabular} & 32 & 48 & 48 & 8 \\ \hline
Processor & \begin{tabular}[c]{@{}l@{}}Intel(R) \\ Xeon(R)\\ CPU E5-2620 v4 \\ @ 2.10GHz\end{tabular} & \begin{tabular}[c]{@{}l@{}}2x 8core\\ Intel \\ E5-2650v2\\  @ 2.6GHz\end{tabular} & \begin{tabular}[c]{@{}l@{}}2x 8core\\ Intel \\ E5-2650v2\\  @ 2.6GHz\end{tabular} & I7-8650U \\ \hline
\begin{tabular}[c]{@{}l@{}}Storage \\ (GB)\end{tabular} & 1024 & 250 & 250 & 8 \\ \hline
\end{tabular}
\label{tab:poc}
\end{table}
\begin{table}[]
\centering
\caption{The mean and standard deviation values for two scenarios.}
\begin{tabular}{|l|l|l|l|}
\hline
\multicolumn{2}{|l|}{\textbf{Scenario}} & \textbf{Mean (ms)} & \textbf{Standard deviation (ms)} \\ \hline
1 & \begin{tabular}[c]{@{}l@{}}No application context \\ relocation\end{tabular} & 331.117 & 117.543 \\ \hline
2 & \begin{tabular}[c]{@{}l@{}}Application context \\ relocation\end{tabular} & 252.924 & 29.786 \\ \hline
\end{tabular}
\label{tab:2}
\end{table}
\subsection{Results}
The application service running on the distributed \gls{mec} hosts in our \gls{poc} are cloud-native Docker-based applications deployed in Kubernetes environment, with RESTful APIs \textcolor{black}{exposed to vehicles for retrieving information about driving conditions on the road in a JSON format}. 

In Fig. \ref{fig:rtt}, we show the trace of the measured \gls{rtt} values for the client running in the vehicle on the Smart Highway, and for all three application servers deployed in distributed \gls{mec} environments, in order to test the impact of the network on the overall service response time, which contains the transmission and propagation delay (network impact), and computational delay on the application server (MEC impact). Furthermore, in Fig. \ref{fig:response}, we show the overall response time of the application server, measured on the client side, for two different scenarios. This response time is important because it shows the delay in retrieving the important contextual driving information from the server, and keeping this response time at a low level (e.g., below 100ms) is essential for vehicle to make decisions. 

In both scenarios, the MEC host 1 is never selected by MEC application orchestrator for an application placement due to the high resource consumption (since we have increased it artificially by performing load stress tests to train our prediction model), while MEC hosts 2 and 3 are being selected based on the projected resource consumption due to the \gls{rtt} of similar scale. \textcolor{black}{In the first scenario no application-context relocation is performed, thus, vehicle remains connected to the \gls{mec} host 2, and as it can be seen in Fig. \ref{fig:response}, once the load increases on the MEC host 2 (after 200s), the response time of the application service is increasing, which means that the driving information about the conditions on the road might be significantly delayed at the vehicle side, leading to the inefficient decisions that will affect the whole manoeuvre experience. On the other hand, in scenario 2, we show that in the case when load increases on the MEC host 2 (i.e., resource availability decreased), as predicted by our algorithm for the time after 200s, the proactive decision on relocating the application-context from application service on the MEC host 2, to MEC host 3, results in the relatively stable response time, which does not increase when vehicle starts retrieving service information from application service on the MEC host 2. Furthermore, a similar decision can be made by our algorithm in case user mobility event notification is received from the core network, and testing such scenario is part of our future work.}

The mean and standard deviation values for both scenarios are shown in Table \ref{tab:2}, and we can see that in scenario 1, when there is no application context relocation for the observations that appear after 200th second, the deviation from the mean is large, i.e., the increase in response time is statistically significant. Thus, in scenario 2, we show that optimized and proactive MEC host selection that results in application-context relocation helps to improve the overall response time, \textcolor{black}{and to prevent service unavailability that leads to outdated information about the conditions on the road, which consequently highly affects the manoeuvre decisions made by vehicle.}

%
%


\section{Conclusion}
In this paper we presented the management and   orchestration   framework   for   vehicular   communications, based on the \gls{3gpp} architecture for enabling edge applications, and \gls{etsi} \gls{nfv}. Such framework enables service continuity for the vehicles by performing an optimized application-context relocation from one edge host to another, thereby allowing vehicle to always connect to the most suitable application server to retrieve the important information about driving conditions on the road. This information is important especially for autonomous vehicles that need to derive decisions about maneuvering without any assistance from the driver. In the experiments on top of the PoC that we created, we show that optimized and proactive application-context relocation helps to improve the overall response time, and to prevent longer delays that cause the outdated information about the conditions on the road. 

\section{Acknowledgement}
This work has been performed in the framework of the H2020 project 5G-CARMEN co-funded by the EU under grant agreement No. 825012. The work has been also supported by the Horizon 2020 Fed4FIRE+ project, Grant Agreement No. 723638. The views expressed are those of the authors and do not necessarily represent the project.



\bibliographystyle{IEEEtran}
\bibliography{IEEEabrv,Bibliography/refs.bib}
%



\end{document}